\documentclass[mathleft
]{an}
\usepackage{graphicx}
\usepackage{times}
\begin{document}

% The following seven commands are intended for editorial usage and should be ignored by
% the author(s).
\Pagespan{1}{4}% Document's page range. 
% If second parameter is left empty, the last page is computed automatically.
\Yearpublication{2008}%
\Yearsubmission{2008}%
\Month{0}%   
\Volume{0}%  
\Issue{0}% 
% \DOI{This.is/not.aDOI}% 

%\title{Astronomische Nachrichten -- \\
%        instructions for authors using \LaTeXe\ markup\,\thanks{Data
%from STELLA}}

\title{High resolution in $z$-direction: \\ The simulation of disc-bulge-halo
galaxies using the particle-mesh code SUPERBOX}

\author{R. Bien \inst{1}\fnmsep\thanks{Corresponding author:
  \email{reinhold@ari.uni-heidelberg.de}\newline},
%Example 
%for footnote, note the usage of the \texttt{fnmsep}
%command as separator between institute number and footnote mark}
A. Just\inst{1}, P. Berczik\inst{1,2}, 
\and I. Berentzen\inst{1}
}
\titlerunning{Improved resolution in $z$ using SUPERBOX}
\authorrunning{R. Bien, A. Just, P. Berczik \& I. Berentzen}
\institute{
Astronomisches Rechen-Institut, Centre for Astronomy (ZAH), University of Heidelberg, \\ 
Moenchhofstr. 12-14, 69120 Heidelberg, Germany
\and
Main Astronomical Observatory, National Academy of Sciences of Ukraine, \\
27 Akademika Zabolotnoho St., 03680 Kyiv, Ukraine
}

\received{30 May 2005}
\accepted{11 Nov 2005}
\publonline{later}

\keywords{galaxies: evolution -- galaxies: kinematics and dynamics -- 
          galaxies: structure -- methods: $N$-body simulations -- methods: numerical}	  

\abstract{%
SUPERBOX is known as a very efficient particle-mesh
code with highly-resolving sub-grids. Nevertheless,
the height of a typical galactic disc is small compared
to the size of the whole system. Consequently, the numerical
resolution in $z$-direction, i.\,e. vertically with respect to the 
plane of the disc,  remains poor. 
Here, we present a new version of SUPERBOX that allows for a 
considerably higher resolution along $z$. The improved code is applied to
investigate disc heating by the infall of a galaxy satellite. 
We describe the improvement and communicate our results. As an 
important application we discuss the disruption of a dwarf 
galaxy within a disc-bulge-halo galaxy that consists of some 
$10^6$ particles.
}

\maketitle

\section{Introduction}

When modelling the dynamics of galaxies, several numerical tech\-ni\-ques 
can be applied. Most useful are direct $N$-body codes, tree codes, and 
particle-mesh codes. 

Two problems, however, become evident. The re\-la\-xa\-ti\-on time
of the stellar disc of a spiral galaxy is larger than the
Hubble time. Therefore, the long-term evolution of a disc
is essentially collisionless. In simulations of unperturbed
galaxies, -- i.e., in the absence of, for instance, stellar
bars, spiral arms, molecular clouds -- the thickness of their
discs is an appropriate measure of how collisionless the
code is: if the unperturbed disc thickens with time, then
two-body encounters prevail and the relaxation time of the
system is unrealistically short. We call this effect
``numerical (or artificial) heating''. (We assume that the number of 
particles is sufficiently large.) As an instructive example, we refer
to Vel{\'a}zquez \& White (1999). They compared their
simulations of sinking satellites (which cause a heating
of the disc) to a corresponding undisturbed model. However,
even in isolation the disc in their models heats up numerically
and thickens. The second concern is the spatial resolution. 
In particular, a code is required to reliably resolve a disc in $z$-direction, 
i.\,e. vertically with respect to the plane of the disc.

SUPERBOX is an improved particle-mesh code and has been applied to a wide variety 
of dynamical problems. Here, we mention
the simulation of the high-velocity encounter of NGC 4782/4783 (Madejsky \& Bien 1993),
the dynamical evolution of a low-mass satellite ga\-la\-xy (Klessen \& Kroupa l998), the 
decay of satellite dwarf ga\-la\-xies in flattened dark matter haloes (Pe\~nar\-ru\-bia et al. 2002), 
dynamical friction in flattened systems (Pe\~narrubia et al. 2004),
and an investigation of large scale inhomogeneity and local dynamical friction 
(Just \& Pe\~nar\-ru\-bia, 2005). Spin\-na\-to et al. (2003) studied the 
inspiral of a black hole to the Galactic centre, and Khoperskov et al. (2007) simulated 
unstable modes in a collisionless disc. Most recently, Fellhauer et al. (2008) published on
the dynamics of the Bootes dwarf spheroidal galaxy, and Pe\~nar\-ru\-bia et al. (2008)
investigated the tidal evolution of dwarf spheroidals of the Local Group.           
    
In this paper, we are aiming to demonstrate that SUPERBOX is both, collisionless 
and capable of high resolution. The main application is the interaction of
a disc-bulge-halo galaxy with an infalling satellite galaxy. 

\section{The philosophy of SUPERBOX}

\begin{figure}
\includegraphics[width=83.6mm]{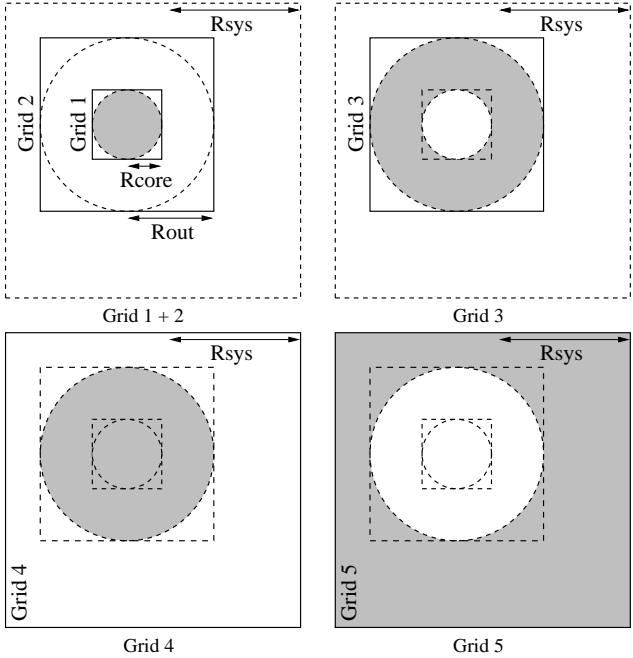}
\caption{The five grids of SUPERBOX.}
\label{label1}
\end{figure}  

SUPERBOX is a particle-mesh code which treats $N$ gravitating 
particles (``superstars'') self-con\-sis\-tent\-ly in a nested system of 3D Cartesian 
grids and sub-grids, so-called boxes, with $n = 2^m$ cells in each dimension. 
In this context, ``self-consistent'' means that, in our models, {\it all} components are represented 
by gravitating particles. Typically, a computation with
$n = 2^7$ and $N = 10^7$ takes a few days on a standard PC or workstation. 
The basic idea behind SUPERBOX is to increase the resolution only at places 
where it is necessary (Bien et al. 1990). For instance, a disc-bulge-halo galaxy 
and an infalling satellite galaxy are placed into a coarse grid (``global grid'', i.e. 
the ``local universe''), while each galaxy belongs to a fi\-ner co-mo\-ving grid, 
called ``outer grid''. Each central region (``core'') is resolved in a yet 
finer sub-grid (``inner grid''). 

More precisely, each galaxy utilizes five grids. In Fig. 1 
these five principal grids are shown. Particles are counted in the shaded areas only. 
$R_\mathrm{core}$ is the radius of the ``core'', $R_\mathrm{out}$ is the maximum 
radius of the outer grid and $R_\mathrm{sys}$ is the radius of the global grid. 
The grids are defined as follows:

\begin{itemize}
\item[--]
{\it Grid 1} is the high-resolution grid and resolves the centre of the galaxy.  
\item[--]
{\it Grid 2} has an intermediate resolution and contains only particles inside $R_\mathrm{core}$.
\item[--]
{\it Grid 3} is equal to grid 2, but it contains only particles between $R_\mathrm{core}$ and 
  $R_\mathrm{out}$.
\item[--]
{\it Grid 4} is the (fixed) global grid and contains only particles inside $R_\mathrm{out}$.
\item[--]
{\it Grid 5} is equal to grid 4, but contains only particles outside $R_\mathrm{out}$. 
\end{itemize}

The corresponding five 
potentials can be linearly superposed to form the total potential. 
Due to the additivity of 
the potential (and hence the accelerations) all galaxies under consideration are treated consecutively 
in the same five grid-arrays. 

\begin{itemize}
\item[--]
For a superstar with $r < R_\mathrm{core}$, the potentials
of grids 1, 3 and 5 are used; $r$ is the distance from the centre.     
\item[--]
For $R_\mathrm{core} < r < R_\mathrm{out}$, the potentials of grids 2, 3, and 5 are used. 
\item[--]
For $r > R_\mathrm{out}$, grids 4 and 5 are used.
\end{itemize}

In principle, the number of galaxies in SUPERBOX can be arbitrary and is
not restricted to two interacting systems (Fellhauer et al. 2000), or an isolated
galaxy.

So far, the resolution of the grids in $z$-direction was not optimal for discs.   
We overcome this problem by flattening grid 2 and grid 3 (the intermediate grids)
along the corresponding $z$ axis when the potential of the first galaxy, i.\,e. 
the disc-bulge-halo galaxy, is 
calculated. Typically, the flattening is $q= 1/4$, and thus the resolution is 
improved by a factor 4. The value of $q$ should not be smaller than 
$$q_\mathrm{crit} = \mathrm{max}\left({R_\mathrm{core} \over R_\mathrm{out}},{4 \over {n-4}}{R_\mathrm{sys} \over R_\mathrm{out}} \right)$$
in order to cover the inner grid and at leat two cells of the global grid in $z$-direction.   

Our present results are discussed in the following.

\section{SUPERBOX vs TREE-GRAPE Code}

\begin{figure}
\includegraphics[width=83.6mm,height=83.6mm]{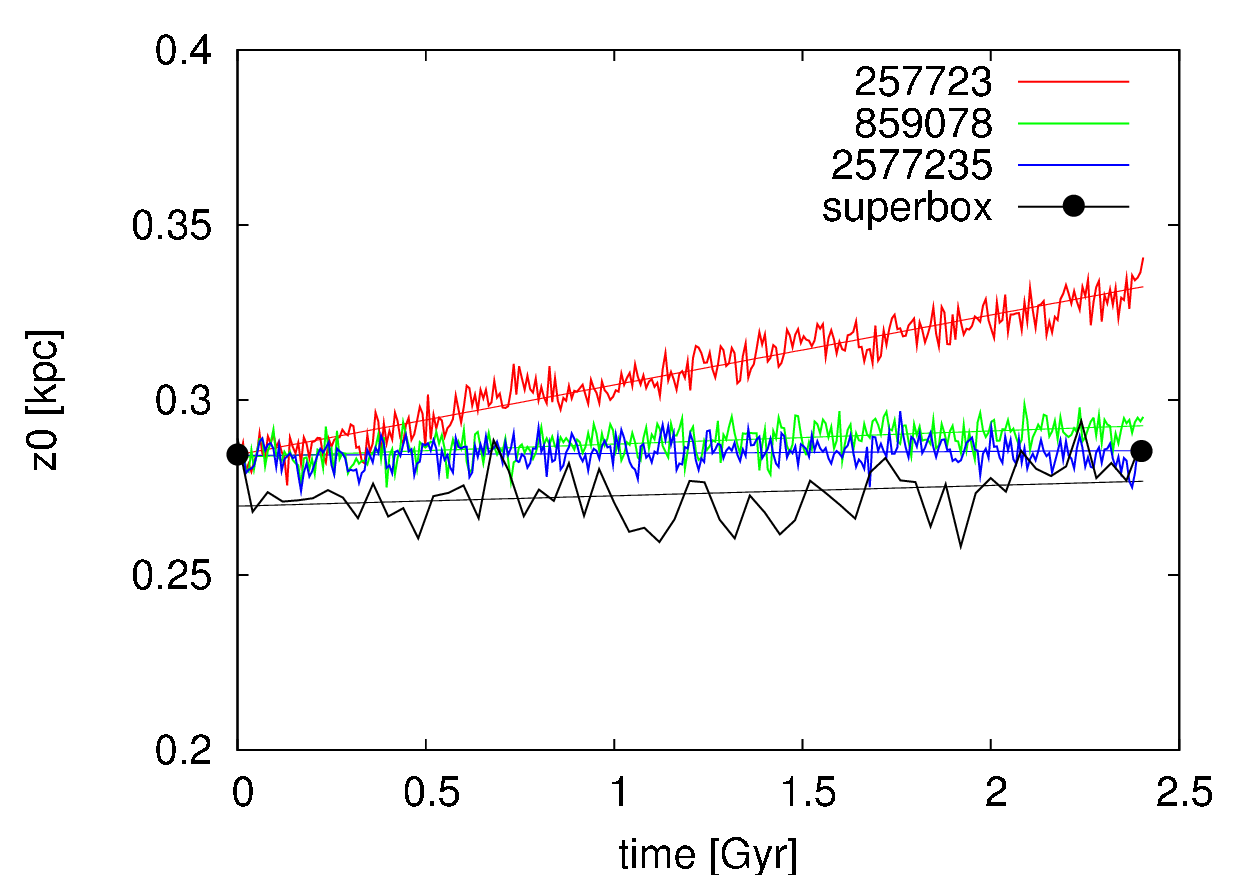}
\caption{The evolution of the scale height $z_\mathrm{0}$.
Shown are the three simulations by the TREE-GRAPE Code and the result found by SUPERBOX.}
\label{label1}
\end{figure} 

We  simulated a disc-bulge-halo galaxy using SUPERBOX. The parameters are

\begin{itemize}
\item[--]
{\it Disc} \\ 190\,000 particles, $2.80 \, 10^{10} \, {\mathrm M}_{\odot}$, 
$R_{\mathrm max} = 35$\, kpc
\item[--]
{\it Bulge} \\ 63\,323 particles, $9.33 \, 10^{9} \, {\mathrm M}_{\odot}$, 
$R_{\mathrm max} = 3.5$\, kpc
\item[--]
{\it Halo} \\ 2\,609\,000 particles, $3.92 \, 10^{11} \, {\mathrm M}_{\odot}$, 
$R_{\mathrm max} = 84$\, kpc
\end{itemize}

We used $R_\mathrm{sys} = 105$\,kpc, $R_\mathrm{out} = 35$\,kpc, 
$R_\mathrm{core} = 3.5$\,kpc, $n = 128$, and $q= 1/4$. Particles outside the global grid are removed from the simulation. 
Thus, the total number of particles decreases very slightly with time. The disc-bulge-halo was built
with the aid of the code MAGALIE, following Boily et al. (2001). Here, it is sufficient to note that 
the disc profile drops exponentially with radius and is isothermal in $z$-direction.

\begin{figure}
\includegraphics[width=83.6mm,height=83.6mm]{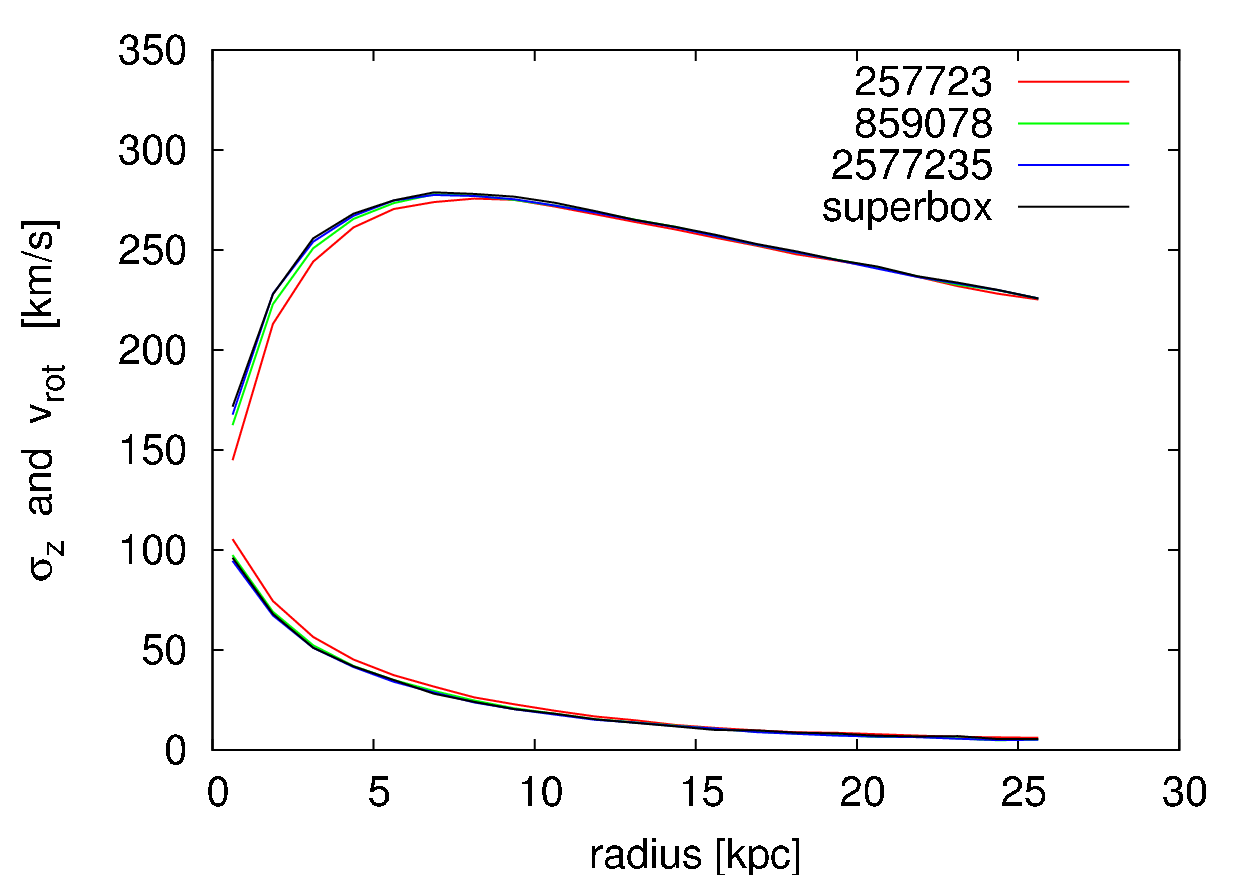}
\caption{Rotation curve $v_{\rm rot}$ and velocity dispersion $\sigma_z$ after 2.4\,Gyr.
Compared are the TREE-GRAPE Code with different numbers of halo particles and SUPERBOX.}
\label{label1}
\end{figure}

A comparison was made with the very popular self-co\-ded TREE-GRAPE sche\-me (Fukushige et al. 2005). 
Such a sche\-me allows us to have a very fast self-gravity calculation routine with up to a few million 
particles per GRAPE node. 
Three simulations with different numbers of halo particles (257\,723, 859\,078, 
and 2\,577\,235) were considered. For the maximum particle number (2\,577\,235) the CPU ti\-me is ab\-out three 
days on the single GRAPE6a board. Likewise, the CPU time of SUPERBOX on a standard PC is of the 
same order. This demonstrates the efficiency of the SUPERBOX code.

Figure 2 shows the scale height of the disc, $z_\mathrm{0}$, versus time; 
$z_\mathrm{0}$ is determined by fitting the vertical density profile 
$$\rho(z) \propto \mathrm{sech}^{2} \left( {z \over {2 z_\mathrm{0}}} \right)$$
 The black colour refers to SUPERBOX results. 
Both dots are calculated using all disc particles for fitting $z_\mathrm{0}$. This demonstrates that the code is intrinsically 
collisionless with negligible numerical heating. The line includes only 15\,\% of the particles. 
The systematic effect of the TREE-GRAPE Code decreases when the number of particles 
increases according to the expected numerical disc heating by massive halo particles.            

Figure 3 shows both, the rotation curve $v_{\rm rot}$ and velocity dispersion $\sigma_z$ of 
the $z$ component after 2.4\,Gyr. As before, the simulations by the TREE-GRAPE code and by SUPERBOX
are compared.  

\section{Example: Highly eccentric intruder}

\begin{figure}
\includegraphics[width=83.6mm]{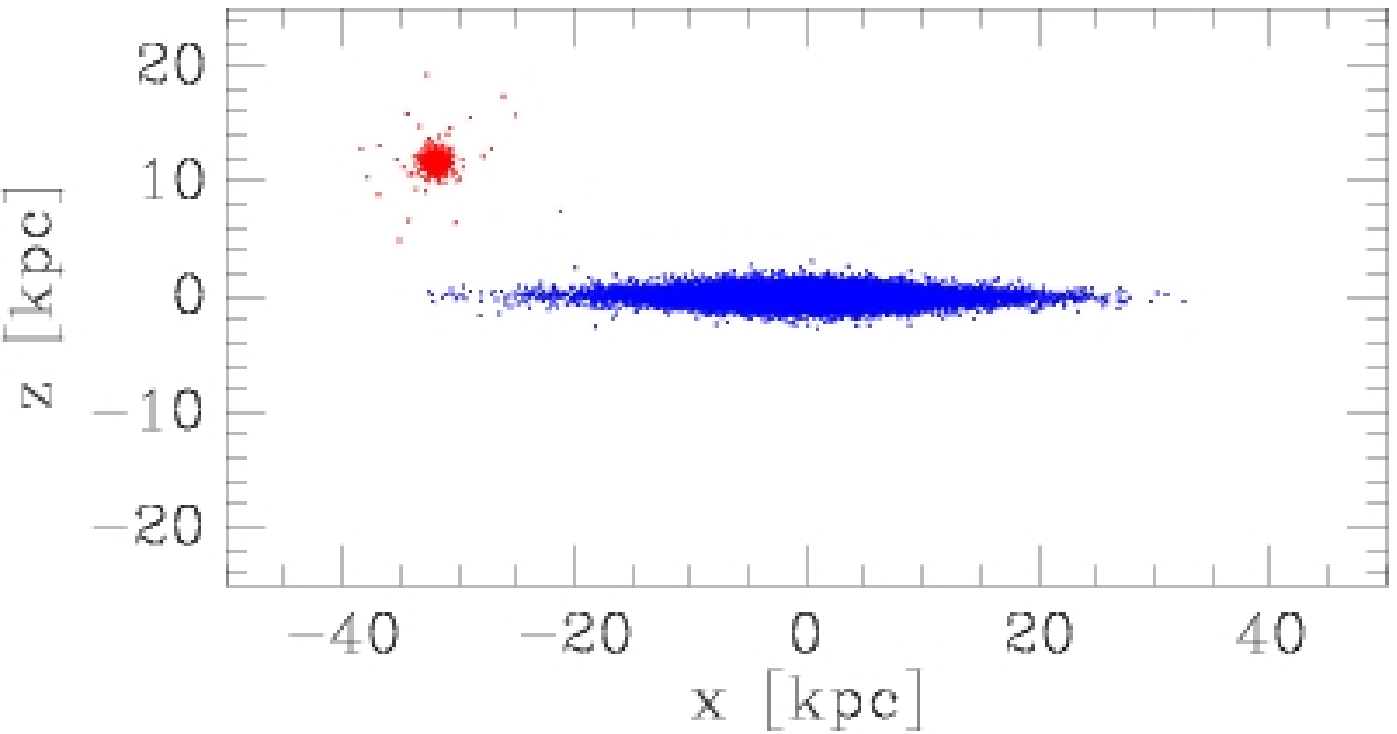}
\includegraphics[width=83.6mm]{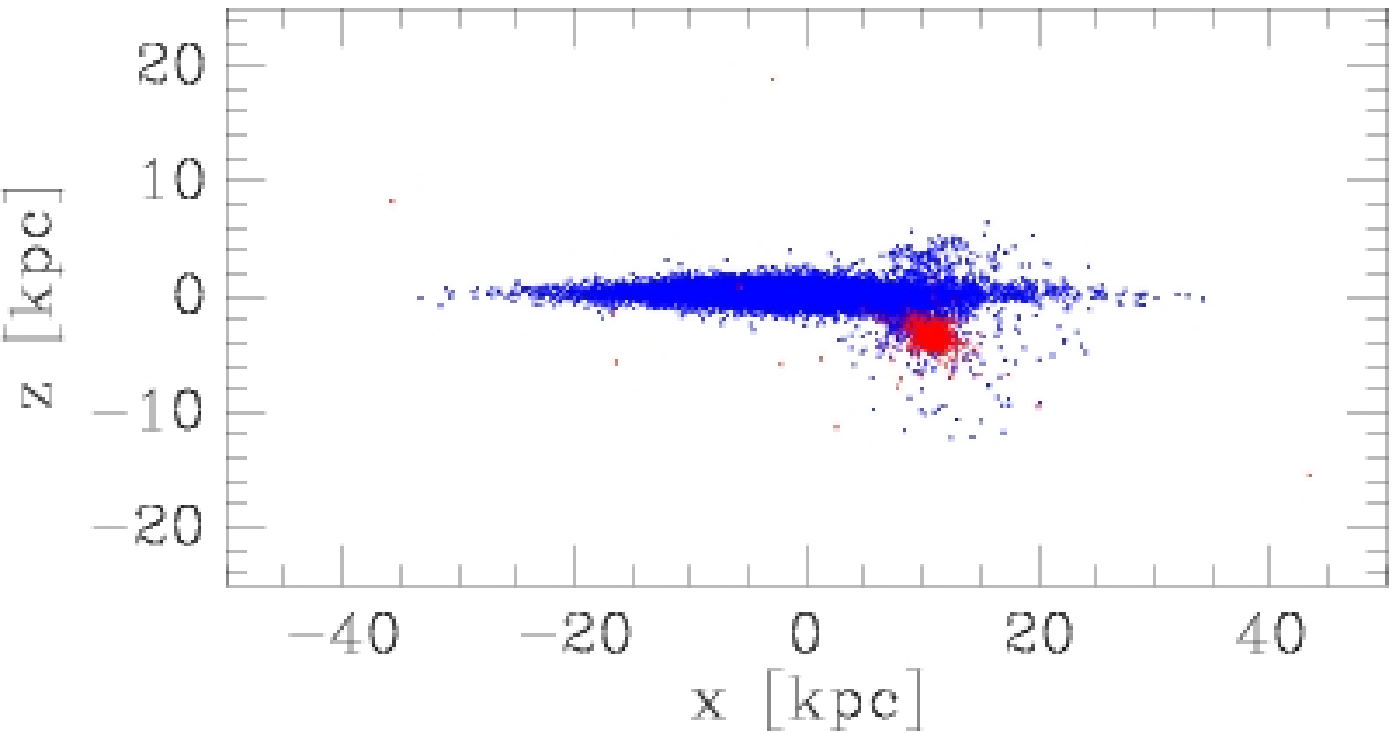}
\includegraphics[width=83.6mm]{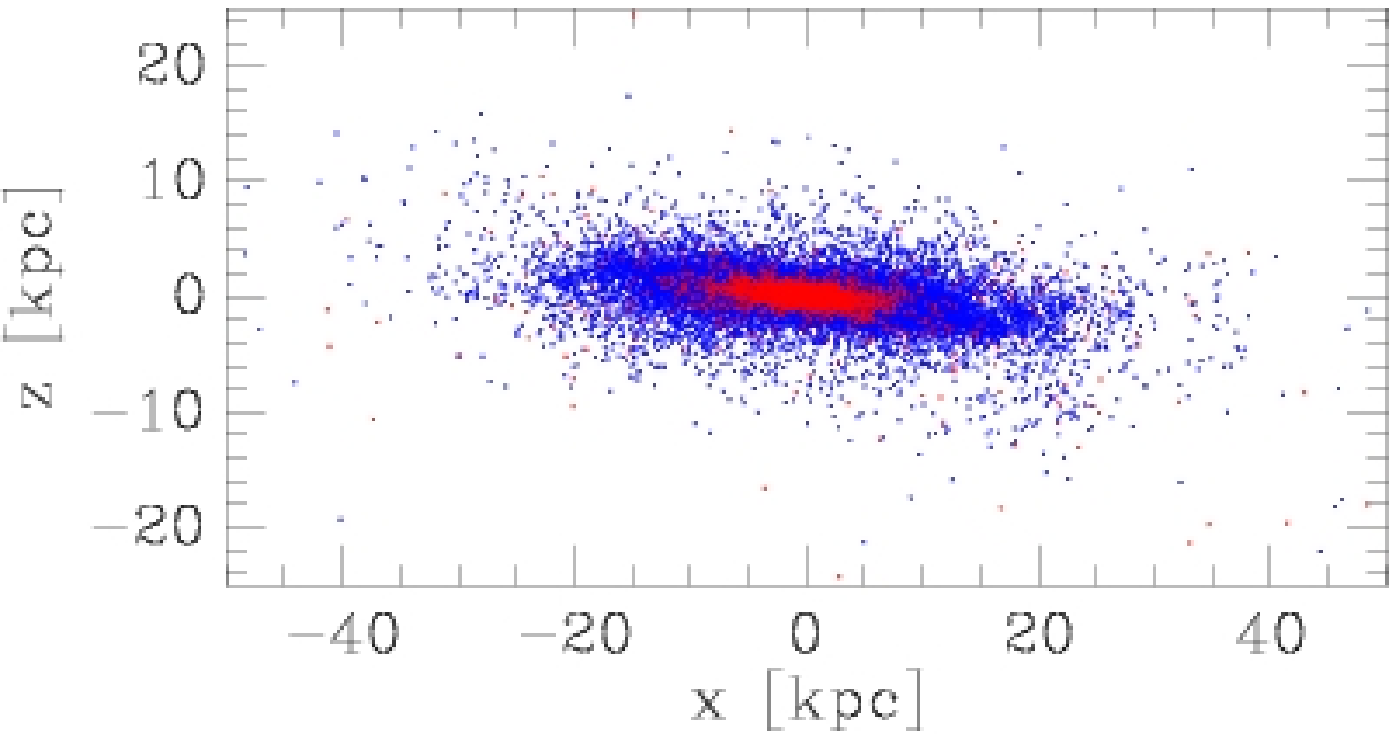}
\caption{A satellite galaxy (red) is colliding with a disc-bulge-halo galaxy (blue). The  
snapshots correspond to 40, 280, and 2\,400\,Myr, from top to bottom.
Bulge and halo particles are not shown.} 
\label{label1}
\end{figure}

In the following, we describe a simulation of the interaction between a disc-bulge-halo galaxy
and a satellite galaxy using SUPERBOX.  
The satellite is represented by a Plummer sphere of 33\,927 particles, corresponding to 
$5 \, 10^{9} \, {\mathrm M}_{\odot}$. It is orbiting on a highly eccentric orbit and 
finally collides with the disc-bulge-halo 
galaxy. Originally, the satellite starts 12\,kpc abo\-ve the plane of the disc 
having a velocity much smaller than the circular speed.
The integration is over 12\,000 time steps, or 2.4\,Gyr. That is, the step size is 0.2\,Myr. 

The snapshots in Fig. 4 show how an inclined and thickened disc 
evolve. Here, only disc particles (blue) and satellite particles (red) are presented. For the sake of clarity, 
bulge and dark halo are not shown. 
The snapshot corresponding to 40\,Myr gives a rough impression 
of the starting conditions. The second snapshot (280\,Myr) shows how the satellite hits the disc. 
An inclined and thickened disc remains 
after 2\,400\,Myr (snapshot 3). In Fig. 5 the 
trajectory of the satellite is projected onto the $x$-$y$ plane.  

\begin{figure}
\includegraphics[width=83.6mm]{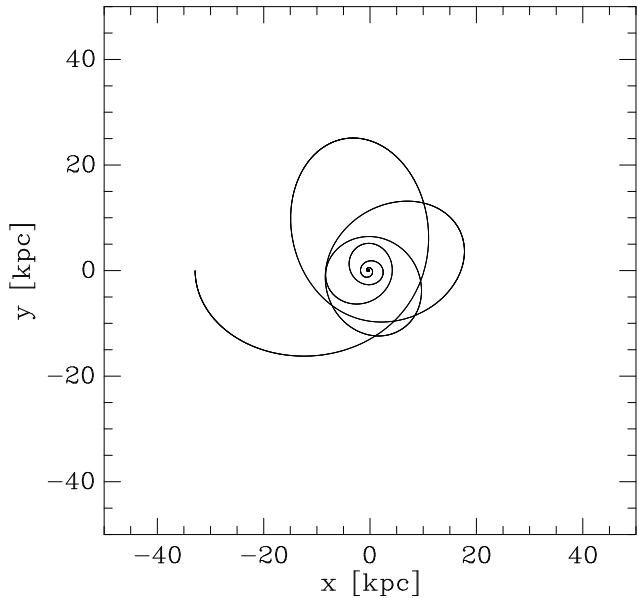}
\caption{The orbit of the satellite is projected onto the $x$-$y$ plane.}
\label{label1}
\end{figure}

\section{Conclusions}

Our preliminary results can be summarized as follows.
  
\begin{itemize}
\item[--]
{\it SUPERBOX is a very fast particle-mesh code}\\ 
This is evident from our comparison with the TREE-GRAPE Code, see sect. 3.  
\item[--]
{\it No numerical disc heating}\\ 
Fig. 2 clearly demonstrates that SUPERBOX is a collisionless
code and thus provides a reliable description of the dynamics of galaxies.
This allows us to study the long-term evolution of disc heating caused by bars,
spiral arms, satellite galaxies, and other mechanisms.     
\item[--]
{\it Improved vertical resolution}\\ 
SUPERBOX has a much better resolution in $z$-direction than before.
It is possible to improve the present flattening up to a factor of 10.
   
\end{itemize}

The new version of SUPERBOX is quite successful and our work on disc heating and 
related effects is in progress. In particular,
use will be made of grids consisting of $2^8 \times 2^8 \times 2^8$ cells
instead of $2^7 \times 2^7 \times 2^7$. We are planning to make the code
publicly available.    
 
\acknowledgements
P. B. acknowledges his support from the German
Science Foundation (DGF) under SFB 439 (sub-project B11)
``Galaxies in the Young Universe'' at the University of
Heidelberg. I. B. acknowledges support from the Volkswagen
Foundation (Volkswagenstiftung) under GRACE Project (Ref. I80 041-043).

\end{document}